\documentstyle[aps,epsf]{revtex}

\begin{document}
%\draft

\title{Effective Drag Between Strongly Inhomogeneous Layers:
Exact Results and Applications}
\author{V. M. Apalkov and M. E. Raikh}
\address{Department of Physics, University of Utah, Salt Lake City,
Utah 84112}

\maketitle

\begin{abstract}
We generalize Dykhne's calculation of the effective resistance of a
2D two-component medium to the case
of frictional drag between the two parallel two-component layers. 
The resulting exact expression for the effective transresistance, 
$\rho^{\mbox{\tiny $D$}}_{\mbox{\small eff}}$,
 is analyzed in the limits
%for the cases
when the resistances and transresistances of the constituting
components are strongly different -  situation generic for
the vicinity of the {\em classical} (percolative) 
metal-insulator transition (MIT).
On the basis of this analysis we conclude that the evolution of
$\rho^{\mbox{\tiny $D$}}_{\mbox{\small eff}}$
%effective drag 
across the MIT is determined by the type
of correlation between the components, constituting the 2D layers.
Depending on this correlation, 
in the case of two electron layers, 
$\rho^{\mbox{\tiny $D$}}_{\mbox{\small eff}}$
changes either monotonically or exhibits a
sharp maximum. For electron-hole layers 
$\rho^{\mbox{\tiny $D$}}_{\mbox{\small eff}}$ is negative and 
$\left| \rho^{\mbox{\tiny $D$}}_{\mbox{\small eff}}\right|$
exhibits a sharp minimum at the MIT. 
\end{abstract}

\pacs{PACS numbers: 71.30.+h, 73.21.Ac,73.40.-c}

%\vspace{7mm}

%\large

\noindent {\em Introduction.} Frictional drag between two
layers has been first predicted
theoretically\cite{pogrebinskii77,price83}
and later observed experimentally\cite{gramila91,sivan92}.
The characteristics measured in experiment is the drag resistance,
$\rho^{\mbox{\tiny $D$}}= V^{p}/I^{a}$, where, $V^{p}$,
is the voltage,  built up in the passive
layer upon passing the current, $I^{a}$, through the active layer.
Experimental observations\cite{gramila91,sivan92}
%observation
have inspired a great number of
theoretical studies of the frictional drag for different
realizations of the 2D electron (hole) systems, constituting  active and passive
layers\cite{tso93,shimshoni94,swierkowski95,vignale96,bonsager96,ussishkin97,sakhi97,shimshoni97,oreg98,gornyi99,khaetskii98,kim99,narozhny00,narozhny01,vonOppen01,felix}.
In parallel, a general formalism for calculating of drag was
advanced\cite{tso92,jauho93,zheng93,flensberg94,kamenev95,flensberg95,flensberg'95}.

In all theoretical papers published by now, the parallel layers were
assumed perfectly homogeneous on the macroscopic scales 
(usually, the scales exceeding the carrier mean free path, $l$). 
Incorporating  tunneling bridges\cite{oreg98} 
or assuming
correlations between the wave functions of the two layers\cite{gornyi99}
 did not violate their {\em macroscopic} homogeneity.
Also, except for
Refs. \onlinecite{narozhny00,narozhny01}, the temperature was assumed
high enough, thus allowing to neglect the mesoscopic fluctuations
due to coherence of different regions of the layers.
The question about the magnitude
of drag between the  2D layers, which are
strongly inhomogeneous {\em macroscopically} was not addressed in 
the theories\cite{tso93,shimshoni94,swierkowski95,vignale96,bonsager96,ussishkin97,sakhi97,shimshoni97,oreg98,gornyi99,khaetskii98,kim99,narozhny00,narozhny01,vonOppen01,felix,tso92,jauho93,zheng93,flensberg94,kamenev95,flensberg95,flensberg'95}.  This question is studied 
in the present paper. 

To be specific, consider first the following situation.
Assume that the passive layer is a good metal, $k_Fl_p \gg 1$,
which is perfectly homogeneous with a fixed concentration of
electrons, $n_p=k_F^2/2\pi$. The concentration, $n_a(\bbox{r})$,
of electrons in the active layer is determined by the concentration 
of donors, $N_D(\bbox{r})$.
Due to, say,  imperfections in the doping process, $N_D(\bbox{r})$
fluctuates on a macroscopic scale with very large correlation length,
$R_c \gg l$ (such an assumption was previously adopted in 
Refs.\onlinecite{simon94,ruzin96}). Assume now, that the active layer
can be depleted by applying the gate voltage, $V_g$. Without the gate
voltage,  $V_g=0$, we have $n_a(\bbox{r})= N_D(\bbox{r})$. 
Upon increasing $V_g$, the electron concentration changes as
$n_a(\bbox{r})= N_D(\bbox{r})-\alpha V_g$, where the dimensionless
coefficient $\alpha$ describes the depletion rate. Assume also, that
at $V_g=0$  the concentration, $n_a$, is high enough, so that 
even with fluctuations,  every region of the active layer is
metallic. As $V_g$ increases, the local resistivities 
$\rho^{a} (\bbox{r}) = 
\rho^{a} \left\{ N_D(\bbox{r})-\alpha V_g\right\}$ will also increase,
but at different rate, so that
within a certain domain of $V_g$ the inhomogeneities 
in $N_D(\bbox{r})$
will become important. Namely, while some regions of the
active layer will remain metallic with $\rho^{a}=\rho^{a}_1$ weakly 
dependent on temperature, the remaining area of the active layer
will turn into insulator with 
$\rho^a=\rho^{a}_2 \propto \exp\left(\mbox{\small ${\cal U}$}/T\right)$, 
where $\mbox{\small ${\cal U}$}$ is the 
activation energy. Then it is clear, that
at certain critical $V_g=V_g^c$, the metallic regions will occupy 
exactly $50\%$ of the area of the active layer. In other words,
the {\em classical} MIT will take place within the narrow
interval $\vert V_g-V_g^c \vert \ll V_g^c$. The width, $\delta V_g$,
of this interval can be related to the critical exponent, 
$t\approx 1.3$,  of  conductivity in the classical 
percolation\cite{stauffer92}.
Indeed, in the limit $\rho^{a}_2 \rightarrow \infty$, the resistivity
near $V_g^c$ diverges as 
$\rho^a(V_g)\sim\rho^{a}_1
\left[\left(V_g^c-V_g\right)/V_g^c\right]^{-t}$. Conversely, in the 
limit $\rho^{a}_1 \rightarrow 0$, we have 
$\rho^a(V_g)\sim\rho^{a}_2
\left[\left(V_g-V_g^c\right)/V_g^c\right]^{t}$.
Then $\delta V_g$ is determined by matching the two behaviors, i.e. 
$\delta V_g/V_g^c = \left(\rho^{a}_1/\rho^{a}_2\right)^{1/2t}$. 
For activated character of transport in insulator, assumed above,
$\delta V_g$ shrinks with temperature as $\exp\left(-\mbox{\small ${\cal U}$}/2tT\right)$.

It is important to note that, in addition to the above 
qualitative picture, there exists a sound {\em quantitative} result
concerning resistivity at 2D classical MIT. 
%much more accurate information about
%$\rho^a(V_g)$ at the classical MIT is available. 
Namely, as it was demonstrated by Dykhne\cite{dykhne1},
the {\em exact} value of $\rho^a$ at $V_g=V_g^c$ is equal to
$\rho^a\! \left(V_g^c\right)=\left(\rho^{a}_1\rho^{a}_2\right)^{1/2}$.
%The behavior of $\rho^a(V_g)$ is depicted schematically in Fig.~1.
Moreover, the product  $\rho^{a}\! \left(V_g^c+ v \right)
\rho^{a}\!\left(V_g^c- v \right)$ is equal to 
$\left(\rho^{a}_1\rho^{a}_2\right)^{1/2}$ for {\em any} $v $. 
Then the question arises about the 
% In the present paper is the 
behavior of the drag resistance, 
$\rho^{\mbox{\tiny $D$}}_{\mbox{\small eff}}$, in the vicinity of the classical MIT. 

%In contrast to the Anderson localization, to which the drag is 
%insensitive as long as the localization radius exceeds relevant 
%microscopic lengths (such as interlayer distance \cite{shimshoni97}), 
%the change of drag across the classical MIT is drastic.
It is obvious that, outside the interval 
$\vert V_g - V_g^c\vert \lesssim 
V_g^c \left(\rho^{a}_1/\rho^{a}_2\right)^{1/2t}$
the effective transresistance 
is equal to $\rho^{\mbox{\tiny $D$}}_1$ on the ``metallic'' side of MIT
and to $\rho^{\mbox{\tiny $D$}}_2$ on the ``insulating'' side,
where $\rho^{\mbox{\tiny $D$}}_{1}$,
$\rho^{\mbox{\tiny $D$}}_{2}$ are  the
transresistivities between the regions with resistance $\rho^{a}_1$
and $\rho^{a}_2$ of the active layer and the metal of the passive layer,
respectively.
This is because, outside the transition
region, the transport is dominated by the current paths
going exclusively through the regions of either low (metallic side)
or high (insulating side) resistance.
The main message of the present paper is that, similar to the
value of  $\rho^a\!\! \left(V_g^c\right)$, the {\em exact} value 
of $\rho^{\mbox{\tiny $D$}}\!\! \left(V_g^c\right)=
\rho^{\mbox{\tiny $D$}}_{\mbox{\small eff}}$ can be found.
%Below we demonstrate that
In particular, in the limit 
$\rho^{\mbox{\tiny $D$}}_{\mbox{\small eff}} \ll \rho^a_{\mbox{\small eff}}$
this value is given by
\begin{equation}
\label{homogeneous}
\rho^{\mbox{\tiny $D$}}_{\mbox{\small eff}}=
             \frac{\rho^{\mbox{\tiny $D$}}_{1}\sqrt{\rho_2^{a}}+
                 \rho^{\mbox{\tiny $D$}}_{2}\sqrt{\rho_1^{a}}}{
                 \sqrt{\rho_1^{a}}+
                 \sqrt{\rho_2^{a}} }.  
\end{equation}
To analyze the temperature dependence of 
$\rho^{\mbox{\tiny $D$}}_{\mbox{\small eff}}$      
one can use for  $\rho^{\mbox{\tiny $D$}}_{1}$ a conventional 
expression for drag between two metals. 
%(see, e.g.  Refs.~\onlinecite{zheng93,kamenev95}). 
Concerning the drag resistivity, $\rho^{\mbox{\tiny $D$}}_{2}$, we have 
assumed that the transport in the insulating regions of the 
active layer is due to activated electrons. For these electrons, 
collisions with electrons in the passive layer, can be viewed
as an additional source of scattering. From here we conclude, 
that both the {\em conductance} and {\em transconductance} for 
the insulating regions are $\propto \exp\left(-\mbox{\small ${\cal U}$}/T\right)$. In 
transresistance, however, this exponent cancels out, so that 
the $T$-dependence of $\rho^{\mbox{\tiny $D$}}_{2}$ is weak.
It is obvious from Eq. (\ref{homogeneous}) that the magnitude 
of $\rho^{\mbox{\tiny $D$}}_{\mbox{\small eff}}$ lies between 
$\rho^{\mbox{\tiny $D$}}_{1}$ and 
$\rho^{\mbox{\tiny $D$}}_{2}\gg\rho^{\mbox{\tiny $D$}}_{1}$.
Since $\rho_1^{a} \ll \rho_2^{a}$, Eq. (\ref{homogeneous}) can 
be simplified to $\rho^{\mbox{\tiny $D$}}_{\mbox{\small eff}}=
\rho^{\mbox{\tiny $D$}}_{1}+ \rho^{\mbox{\tiny $D$}}_{2}
\left[\rho_1^{a}/\rho_2^{a}\right]^{1/2}$, so that at low $T$ we have
$\rho^{\mbox{\tiny $D$}}_{\mbox{\small eff}} \propto T^2$. 
With increasing $T$ this dependence crosses over to
$\rho^{\mbox{\tiny $D$}}_{\mbox{\small eff}} \propto 
\exp\left(-\mbox{\small ${\cal U}$}/2T\right)$, i.e. becomes activational.
From Eq. (\ref{homogeneous}) we also  conclude that the effective
drag does not follow the evolution of resistivity, 
$\rho^{a}_{\mbox{\small eff}}$, as the classical MIT is continuously 
swept,  due to  the variation of the gate voltage.
Indeed,  the $\rho^{a}_{\mbox{\small eff}}$
changes sharply from $\rho_1^{a}$ on the metallic side
to  $\left(\rho_1^{a}\rho_2^{a}\right)^{1/2}$ at the percolation
threshold, and further to
$\rho_2^{a}$ on the insulating side. On the other hand, the crossover
of $\rho^{\mbox{\tiny $D$}}_{\mbox{\small eff}}$ from 
$\rho^{\mbox{\tiny $D$}}_{1}$ to $\rho^{\mbox{\tiny $D$}}_{2}$ is
``delayed'', as illustrated in Fig. 1. 

The reason why the exact expression 
for $\rho^{\mbox{\tiny $D$}}_{\mbox{\small eff}}$ can be obtained
is that the duality transformation\cite{dykhne1}, which yields a
closed equation for the $\rho^a_{\mbox{\small eff}}$ at MIT 
can be generalized to the
case of two layers. This is because, the double-layer system can 
be viewed as a two-component system, in which
each component consists of {\em two} vertically-separated islands, 
coupled by the mutual drag. Duality transformation between 
the two types  of coupled islands, as depicted in Fig.~1, renders a 
closed {\em matrix} 
equation for the components of the effective resistivity
matrix of the two-layer system. The corresponding steps are 
outlined below.

\vspace{3mm}

\noindent {\em Derivation.} In the presence of drag, each
component of the double-layer system  is characterized by its 
resistivity matrix
\begin{equation}
\label{R1}
\bbox{{\cal R}}_1 = \left( \begin{array}{cc}
                 \rho_1^a & \rho^{\mbox{\tiny $D$}}_{1}
                                                      \\
                  \rho^{\mbox{\tiny $D$}}_{1}   & \rho^p_{1}
                \end{array} \right) ~~,~~
%\end{equation}
%and
%\begin{equation}
%\label{R2}
\bbox{{\cal R}}_2 = \left( \begin{array}{cc} \rho^a_2 &
                     \rho^{\mbox{\tiny $D$}}_{2}
                                                        \\
                  \rho^{\mbox{\tiny $D$}}_2   & \rho^p_{2}
                 \end{array} \right)  .
\end{equation}
If the two components are equally distributed
over the plane, then the effective resistivity matrix,
$\bbox{{\cal R}}_{\mbox{\small eff}}$, can be found
exactly. As we demonstrate below the corresponding expression for 
$\bbox{{\cal R}}_{\mbox{\small eff}}$ has the form
\begin{eqnarray}
\bbox{{\cal R}}_{\mbox{\small eff}} & = & \left( D_1 D_2 \right)^{1/4}
  \frac{ D_2^{1/2} \bbox{{\cal R}}_1+D_1^{1/2} \bbox{{\cal R}}_2 }
    {\sqrt{\det\left[D_2^{1/2} \bbox{{\cal R}}_1+D_1^{1/2} 
\bbox{{\cal R}}_2 \right] } }
           \nonumber \\
        & = & \frac{D_2^{1/2} \bbox{{\cal R}}_1+D_1^{1/2} \bbox{{\cal R}}_2}
      {\sqrt{\left( D_1^{1/2}+ D_2^{1/2} \right)^2 -
      \det\left(\bbox{{\cal R}}_1 - \bbox{{\cal R}}_2\right) } }  ,
\label{main}
\end{eqnarray}
where $D_1 = \det (\bbox{{\cal R}}_1)$ and  $D_2 = 
\det (\bbox{{\cal R}}_2)$ are the determinants of the matrices 
$\bbox{{\cal R}}_1$ and $\bbox{{\cal R}}_2$, respectively.

In general, the calculation of the effective resistivity
requires the solution of the local Ohm equations
\begin{eqnarray}
\label{ohm1}
& &   \bbox{E}_a =  \rho^a \bbox{J}_a
  +  \rho^{\mbox{\tiny $D$}} \bbox{J}_p \\
& &                         \bbox{E}_p =  \rho^p \bbox{J}_p
  +  \rho^{\mbox{\tiny $D$}} \bbox{J}_a
\label{ohm2}
\end{eqnarray}
within each double-layer island constituting one of the
two components, see Fig.~1.
Naturally, Eqs.~(\ref{ohm1}), (\ref{ohm2}) imply the
in-plane isotropy of each component. Then it is convenient
to view the pairs $(\bbox{J}_a,\bbox{J}_p)$ and
$(\bbox{E}_a,\bbox{E}_p)$ as two-component vectors
\begin{equation}
 \hat{\bbox{J}} = \left(\begin{array}{c} \bbox{J}_a \\ \bbox{J}_p
           \end{array}
 \right)  ~~~,~~~  \hat{\bbox{E}}=
                \left(\begin{array}{c} \bbox{E}_a \\ \bbox{E}_p
           \end{array}
 \right),
\end{equation}
and rewrite local equations (\ref{ohm1}), (\ref{ohm2}) in the form
\begin{equation}
\label{ohm}
\hat{\bbox{E}}
= \bbox{{\cal R}}\hat{\bbox{J}},
\end{equation}
where the matrix, $\bbox{{\cal R}}$, assumes one of 
the forms (\ref{R1}) within each component.

Local equations (\ref{ohm}) should be solved together with
Maxwell's and continuity equations
\begin{equation}
\left[ \bbox{\nabla}\times  \hat{\bbox{E}} \right] =
\left( \begin{array}{c}
   \left[ \bbox{\nabla}\times  \bbox{E}_a \right] \\
   \left[ \bbox{\nabla}\times  \bbox{E}_p \right]
       \end{array}
\right) = 0 ~~~,~~~
  \left( \bbox{\nabla} \hat{\bbox{J}}\right) =
  \left( \begin{array}{c}
    \bbox{\nabla} \bbox{J}_a  \\
    \bbox{\nabla} \bbox{J}_p
 \end{array}
\right) = 0  .
 \label{maxwell}
\end{equation}

In order to derive Eq.~(\ref{main}) we demonstrate that, for
globally equivalent distributions of the two components, the
matrix  $\bbox{{\cal R}}_{\mbox{\small eff}}$ satisfies
the following equation
\begin{equation}
 \bbox{{\cal R}}_{\mbox{\small eff}} =
\bbox{{\cal R}}_1\bbox{{\cal R}}_{\mbox{\small eff}}^{-1}
     \bbox{{\cal R}}_2  .
 \label{Geff2}
\end{equation}
This equation generalizes the Dykhne result\cite{dykhne1}
to the case of two layers coupled by drag. It is easy to
see that in the absence of drag, when the matrices $\bbox{{\cal R}}_1$
and $\bbox{{\cal R}}_2$ are diagonal, Eq.~(\ref{Geff2}) immediately
yields the conventional expressions $\rho^{a}_{\mbox{\small eff}}=
\sqrt{\rho^{a}_1\rho^{a}_2  }$ and $\rho^{p}_{\mbox{\small eff}}=
\sqrt{\rho^{p}_1\rho^{p}_2  }$.
In deriving the closed equation (\ref{Geff2}) for
$ \bbox{{\cal R}}_{\mbox{\small eff}}$ we follow  the line
of reasoning put forward by Dykhne\cite{dykhne1}.
Namely, along with $\hat{\bbox{J}}$ and $\hat{\bbox{E}}$,
we introduce the auxiliary variables $\hat{\bbox{J}}_d $ and
$\hat{\bbox{E}}_d $, defined as
\begin{equation}
 \hat{\bbox{J}}_d = \bbox{A}_J
   \left[ \bbox{n}\times \hat{\bbox{E}} \right]
 ~~~~,~~~~
 \hat{\bbox{E}}_{d} = \bbox{A}_E
     \left[ \bbox{n}\times \hat{\bbox{J}}
          \right],
\end{equation}
where $ \bbox{A}_J $ and $ \bbox{A}_E $ are some {\em constant}
matrices, and $\bbox{n}$ is the unit vector normal to the
layers.
It is easy to check that, similarly to $\hat{\bbox{E}}$ and
$\hat{\bbox{J}}$,
the  variables, $ \hat{\bbox{E}}_d$ and
 $\hat{\bbox{J}}_{d}$ also satisfy the Maxwell and the continuity
equations
\begin{equation}
\left[ \bbox{\nabla}\times  \hat{\bbox{E}}_d  \right] = 0 ~~~,~~~
   \left( \bbox{\nabla} \hat{\bbox{J}}_d \right) = 0.
\end{equation}
On the other hand, the Ohm's law Eq. (\ref{ohm}) dictates the following
relation between $\hat{\bbox{E}}_d$ and $\hat{\bbox{J}}_{d}$
\begin{equation}
\hat{\bbox{E}}_d = \bbox{A}_E \left[ \bbox{n}\times \hat{\bbox{J}}
                   \right] =
  \bbox{A}_E\left[ \bbox{n}\times \left(\bbox{{\cal R}}^{-1}
\hat{\bbox{E}}\right) \right] =\left(
\bbox{A}_E\bbox{{\cal R}}^{-1}\bbox{A}_J^{-1}\right) \hat{\bbox{J}}_d
= \bbox{{\cal R}}_d \hat{\bbox{J}}_d.
\label{dual}
\end{equation}
At this point we impose the duality conditions. Namely, we
require that within the {\em first} component $\hat{\bbox{J}}_d$ and
$\hat{\bbox{E}}_d$ are related via the matrix $\bbox{{\cal R}}_2$,
and, conversely, within the {\em second} component the relation
$\hat{\bbox{J}}_d=\bbox{{\cal R}}_1\hat{\bbox{E}}_d$ holds.
If these conditions are met, then the equivalent distribution of the
first and second components guarantees that {\em on average}
$\hat{\bbox{J}}_d$ and $\hat{\bbox{E}}_d$ are related by the same
effective resistivity matrix $\bbox{{\cal R}}_{\mbox{\small eff}}$
as the {\em  average} vectors  $\hat{\bbox{J}}$ and $\hat{\bbox{E}}$.
Quantitatively, the duality conditions are expressed as
\begin{equation}
 \bbox{{\cal R}}_1 = \bbox{A}_E\bbox{{\cal R}}_2^{-1}
 \bbox{A}_J^{-1}    ~~~~,~~~~
 \bbox{{\cal R}}_2 = \bbox{A}_E\bbox{{\cal R}}_1^{-1} \bbox{A}_J^{-1}.
 \label{G12}
\end{equation}
It is easy to see that these conditions are satisfied by choosing
\begin{equation}
\bbox{A}_E=\bbox{{\cal R}}_1 ~~~,~~~
        \bbox{A}_J = \bbox{{\cal R}}_2^{-1}.
\label{AB}
\end{equation}
As a final step, Eq. (\ref{Geff2}) emerges from the following
chain of identities for average fields and currents
\begin{equation}
\langle \hat{\bbox{E}}_d \rangle=
\bbox{A}_E \left[ \bbox{n}\times \langle \hat{\bbox{J}}
           \rangle
                   \right] =
  \bbox{A}_E\left[ \bbox{n}\times
       \left(\bbox{{\cal R}}_{\mbox{\small eff}}^{-1}
\langle \hat{\bbox{E}}\rangle \right) \right] =
\left(
\bbox{A}_E\bbox{{\cal R}}_{\mbox{\small eff}}^{-1}
      \bbox{A}_J^{-1}\right)
\langle \hat{\bbox{J}}_d  \rangle
= \bbox{{\cal R}}_{\mbox{\small eff}}~\! \langle \hat{\bbox{J}}_d
        \rangle .
\label{dual_av}
\end{equation}
With $\bbox{A}_E = \bbox{{\cal R}}_1$ and
$\bbox{A}_J^{-1}=\bbox{{\cal R}}_2$, the last identity in
Eq. (\ref{dual_av}) yields Eq. (\ref{Geff2}).
In general, the effective resistivity matrix is symmetric, and, thus, is
characterized by three unknown elements. As a result, Eq. (\ref{Geff2})
can be reduced to three second-order algebraic equations. It turns
our that only two of them are independent. More precisely,
the general solution of Eq.~(\ref{Geff2}) can be presented in the form
\begin{equation}
\bbox{{\cal R}}_{\mbox{\small eff}} = \alpha \bbox{{\cal R}}_1 +
 \beta \bbox{{\cal R}}_2  ,
 \label{Geff4}
\end{equation}
where $\alpha $ and $\beta $ are the {\em numbers}. In order to
find these numbers, it is sufficient to derive two relations between
them. The first relation expresses the fact that the determinants
of the l.h.s. and r.h.s. of Eq.~(\ref{Geff2}) are equal. This yields
\begin{equation}
\label{det}
\det \left(  \alpha \bbox{{\cal R}}_1 +
 \beta \bbox{{\cal R}}_2   \right)  = \left( D_1 D_2 \right)^{1/2}.
\end{equation}
The second relation emerges upon direct substitution of
Eq.~(\ref{Geff4}) into (\ref{Geff2}) leading to
\begin{equation}
\alpha^2  \bbox{{\cal R}}_1  \bbox{{\cal R}}_2^{-1} +
         \beta^2  \bbox{{\cal R}}_2
  \bbox{{\cal R}}_1^{-1}  =
\left(1-2 \alpha \beta\right)\bbox{{\cal I}} ,
\label{ab1}
\end{equation}
where $\bbox{{\cal I}}$ is the unity matrix. It follows from the above
relation that nondiagonal elements of the l.h.s. are zero, so that
\begin{equation}
\label{ratio}
\frac{\alpha ^2}{\beta ^2} =
\frac{\det(\bbox{{\cal R}}_2)}{\det(\bbox{{\cal R}}_1)} =
\frac{D_2}{D_1}.
\end{equation}
Upon solving the system Eqs. (\ref{det}) and (\ref{ratio}), we find
the following expressions for coefficients $\alpha$ and $\beta$
\begin{equation}
\alpha = \frac{ D_1^{1/4} D_2^{3/4} }
       {\sqrt{\det\left[D_2^{1/2} \bbox{{\cal R}}_1+D_1^{1/2}
        \bbox{{\cal R}}_2 \right] } } ~~,~~
\beta = \frac{ D_2^{1/4} D_1^{3/4} }
       {\sqrt{\det\left[D_2^{1/2} \bbox{{\cal R}}_1+D_1^{1/2}
        \bbox{{\cal R}}_2 \right] } }.
\label{C}
\end{equation}
Substituting these expressions into Eq. (\ref{Geff4}), we arrive at the 
explicit form Eq. (\ref{main}) of the effective resistivity matrix.

\vspace{3mm}

\noindent{\em Applications}. In all realistic situations 
the drag-related nondiagonal components of the matrices  (\ref{R1})
%, being responsible for drag, 
are much smaller than the diagonal components, which describe the
in-plane transport. Under this condition, the effective drag resistance
between the 2D layers  can be simplified to
\begin{equation}
\label{rho_D}
\rho^{\mbox{\tiny $D$}} _{\mbox{\small eff}}=
\frac{\rho^{\mbox{\tiny $D$}}_1 \sqrt{\rho^a_2\rho^p_2} +
\rho^{\mbox{\tiny $D$}}_2 \sqrt{\rho^a_1\rho^p_1}}
{\sqrt{\rho^a_1\rho^p_2} +
 \sqrt{\rho^a_2\rho^p_1} }.
\end{equation}
The case of drag between a homogeneous layer and a two-component system,
considered in the Introduction, corresponds to $\rho^p_2 = \rho^p_1$.
Then Eq. (\ref{rho_D}) immediately reduces to Eq. (\ref{homogeneous}).
Below we consider two more realizations of the double-layer system,
in which both layers are strongly inhomogeneous.

\noindent{\em (i) Symmetric layers.} This situation (see Fig.~2) emerges when
both layers are identical (e.g., positioned symmetrically with respect
to the donors). Moreover, we will assume for simplicity
that the gate voltages applied to the both layers are the same.
Then, in the vicinity of the classical MIT, the islands (see Fig. 1)
will be composed of either two metallic or two insulating components.
Substituting $\rho^p_1 = \rho^a_1$ and  $\rho^p_2 = \rho^a_2$ into
Eq. (\ref{rho_D}) we obtain
\begin{equation}
\rho^{\mbox{\tiny $D$}}_{\mbox{\small eff}} =
\frac{1}{2} \left[ \rho^{\mbox{\tiny $D$}}_{2}  
   \left( \frac{\rho^a_1}{\rho^a_2} \right)^{1/2} \!\! \! \! \!\!+
                   \rho^{\mbox{\tiny $D$}}_{1}  
   \left( \frac{\rho^a_1}{\rho^a_2} \right)^{-1/2}
  \right].
%\frac{\rho^a_1\rho^{\mbox{\tiny $D$}}_{2}+
%          \rho^a_2\rho^{\mbox{\tiny $D$}}_{1}}
%  {2\sqrt{\rho^a_1\rho^a_2}} 
\label{case2_1}
\end{equation}
In contrast to Eq.~(\ref{homogeneous}),  $~ \rho^{\mbox{\tiny $D$}}_{1}$
and $\rho^{\mbox{\tiny $D$}}_{2}$ now stand for transresistances between
two metals and two insulators. Similar to the case of a homogeneous
passive layer, {\em outside} the MIT region we have
$\rho^{\mbox{\tiny $D$}}_{\mbox{\small eff}}=
\rho^{\mbox{\tiny $D$}}_{1}$ and
$\rho^{\mbox{\tiny $D$}}_{\mbox{\small eff}}=
\rho^{\mbox{\tiny $D$}}_{2}$, respectively.
% (Fig. 2). 
However, the
behavior of $\rho^{\mbox{\tiny $D$}}_{\mbox{\small eff}}$ within
the transition region is drastically different from that in Fig.~1.
Indeed, the first term in Eq. (\ref{case2_1}) contains a small factor 
$\left[\rho^a_1/\rho^a_2\right]^{1/2}\propto \exp\left(-\mbox{\small ${\cal U}$}/2T\right)$,
while the second term contains a large factor $\propto\exp\left(\mbox{\small ${\cal U}$}/2T\right)$.
Thus, despite $\rho^{\mbox{\tiny $D$}}_{2} \gg \rho^{\mbox{\tiny $D$}}_{1}$,
at low temperatures the second term will not only dominate, but 
can exceed $\rho^{\mbox{\tiny $D$}}_{2}$. As a result, 
$\rho^{\mbox{\tiny $D$}}_{\mbox{\small eff}}$ will exhibit a 
maximum as a function of $V_g$ in the vicinity of MIT, as 
illustrated in Fig. 2.

\noindent{\em (ii) Electron-hole layers.} The sign of transresistance
 in this case is  negative\cite{sivan92}. 
The phenomenon of drag in the system of {\em homogeneous} 
electron-hole layers was previously
considered in Refs. \onlinecite{tso93,swierkowski95,vignale96}
with an emphasis on the role of interaction-induced correlations
between electrons and holes beyond the random-phase approximation.
We will consider the spatially inhomogeneous situation, assuming that,
without disorder, the concentrations of electrons and holes are
strictly equal. We will also assume that the disorder potential, acting on electrons
and holes, is the same. The crucial observation is that, due to
their opposite charges, electrons and holes ``react'' differently
to the disorder potential. The same potential that creates a ``metallic
lake'' of electrons would deplete the corresponding passive region
of holes, turning them into insulator. As a result, as the MIT
is approached, we arrive to the situation, depicted in Fig. 3,
when the islands consist of pairs of metallic electrons and
insulating holes and vice versa.
Then, substituting  
$\rho^{\mbox{\tiny $D$}}_{2}=\rho^{\mbox{\tiny $D$}}_{1}$ ,
$\rho^p_1 = \rho^a_2$, and  $\rho^p_2 = \rho^a_1$ into 
Eq. (\ref{rho_D}) we get
\begin{equation}
\label{case3_1}
\rho^{\mbox{\tiny $D$}} _{\mbox{\small eff}}=
 - 2 \left\vert \rho^{\mbox{\tiny $D$}}_1 \right\vert
\frac{\sqrt{\rho^a_1\rho^a_2}}
{\rho^a_1+\rho^a_2} .
\end{equation}
It is obvious from Eq. (\ref{case3_1}) that, since
$ \vert \rho^{\mbox{\tiny $D$}}_{\mbox{\small eff}} \vert\sim 
 \left\vert \rho^{\mbox{\tiny $D$}}_1 \right\vert  \exp \left(-\mbox{\small ${\cal U}$}/2T \right)$,
the absolute value of the effective drag exhibits a {\em minimum} 
near $V_g^c$, as illustrated in Fig. 3.

\vspace{3mm}

\noindent{\em Discussion}. In addition to the drag resistance
at MIT, Eq. (\ref{main}) allows to calculate the drag-induced
corrections to the effective conductivity of individual layers.
For the case of a homogeneous passive layer, considered in the
Introduction, this correction has a form
\begin{equation}
\label{correction}
\sigma _{\mbox{\small eff}} -
\frac{1}{\sqrt{\rho^a_1 \rho^a_2}}=
 -  \frac{1}{2\rho^p\sqrt{\rho^a_1 \rho^a_2 }}
\left(\frac{\rho^{\mbox{\tiny $D$}}_1 -
       \rho^{\mbox{\tiny $D$}}_2}{\sqrt{ \rho^a_1}+
\sqrt{ \rho^a_2}}
 \right)^2  .
\end{equation}
It is noteworthy, that the sign of the correction Eq.~(\ref{correction})
is strictly {\em negative}. The reason for that is the underlying
physics of the drag phenomenon. Namely, for electrons in a
passive layer, their  interaction with electrons in an active layer,
that gives rise to drag,  can be also viewed as an
additional source of their scattering. Therefore at the special point
$\rho^{\mbox{\tiny $D$}}_1=\rho^{\mbox{\tiny $D$}}_2$, 
when the r.h.s. of
Eq.~(\ref{correction}) turns to zero, it should be expected that the
fourth-order correction to  $\sigma _{\mbox{\small eff}}$, neglected
in Eq.~(\ref{correction}), has also a negative sign.

Physical explanation of the fact that
$\vert\rho^{\mbox{\tiny $D$}}_{\mbox{\small eff}}\vert$ between the
electron-hole layers has a minimum at MIT is straightforward. Indeed,
when metallic lakes of electrons are located opposite to the
insulating regions of holes (see Fig. 3), then, at MIT,
the current paths in the active layer are perpendicular to those
in the passive layer, so that the conditions for drag are unfavorable.
The origin of maximum of $\rho^{\mbox{\tiny $D$}}_{\mbox{\small eff}}$
at MIT for two correlated electron layers, as  depicted in Fig. 2, is
less transparent. One can speculate that the maximum is due to
the fact that, at MIT, the current paths in two layers are long,
and that, due to perfect correlation, each long path in the active
layer has its ``counterpart'' in the passive layer.
Note finally, that Eq. (\ref{main}) is exact and takes into account 
{\em all } the orders in $\rho^{\mbox{\tiny $D$}}$. 
Although modeling of the classical MIT with two-component mixture
is crude, we believe that, due to strong
difference in resistances of the components, our predictions 
(\ref{homogeneous}), (\ref{case2_1}), and (\ref{case3_1})
for different types of behavior of $\rho^{\mbox{\tiny $D$}}_{\mbox{\small eff}}$
 across the MIT remain valid for realistic 
situations. 

\vspace{3mm}

\noindent{\em Acknowledgements. } One of 
the authors (M.E.R.) is grateful
to the  Weizmann Institute of Science for hospitality, and especially 
to F. von Oppen and A. Stern for highly illuminating discussions.
% on drag during his visit. 
The work was supported by the NSF under Grant No. INT-0231010.

\begin{figure}
\centerline{
\epsfxsize=3.8in
\epsfbox{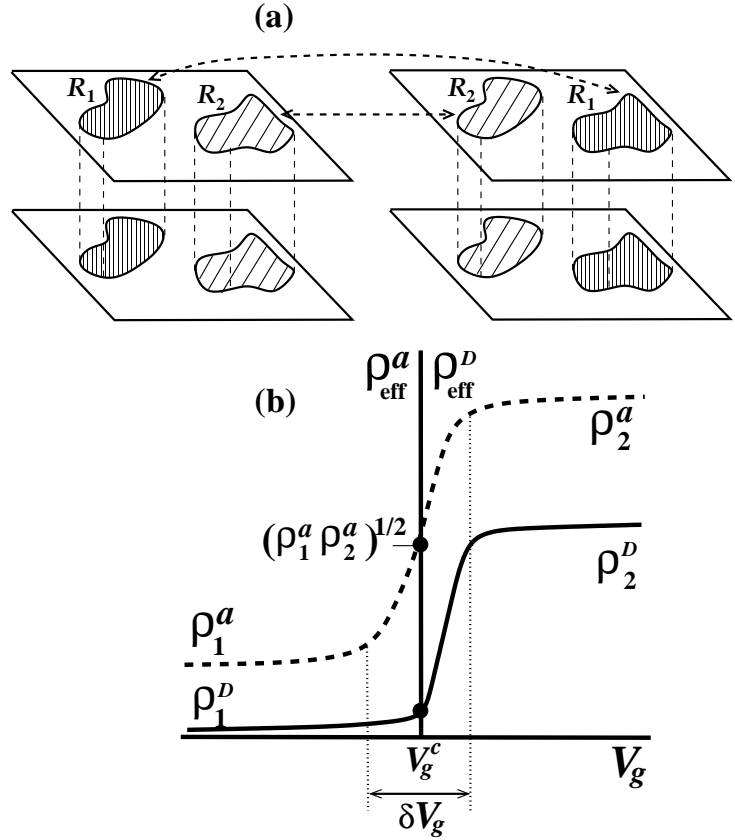}
}
\vspace*{0.1in}
\protect\caption[sample]
{\sloppy{(a) Schematic illustration of the {\em matrix} duality 
transformation. Mutually dual two-layer islands are connected by
horizontal lines; (b) The  resistivity of the active layer 
(dashed line) and the effective drag (solid line) are depicted
as a function of the gate voltage for the case when the passive layer
is a homogeneous metal. The value 
$\rho^{\mbox{\tiny $D$}}_{\mbox{\small eff}}$ at MIT is given by 
Eq. (\ref{homogeneous}), so that 
$\rho^{\mbox{\tiny $D$}}_{\mbox{\small eff}}/
\rho^{\mbox{\tiny $D$}}_{2} \approx (\rho^a_1/\rho^a_2)^{1/2}\ll 1$.
  }}
\label{figone}
\end{figure}

\begin{figure}
\centerline{
\epsfxsize=3.5in
\epsfbox{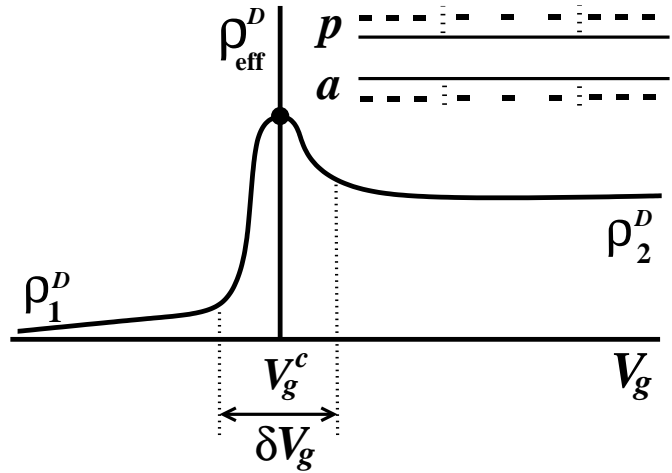}
}
\vspace*{0.1in}
\protect\caption[sample]
{\sloppy{The transresistance across the MIT is depicted schematically
for two correlated electron layers at low $T$. The value
$\rho^{\mbox{\tiny $D$}}_{\mbox{\small eff}}$ at MIT is given by
Eq.  (\ref{case2_1}). The dependence $\rho^a_{\mbox{\small eff}}(V_g)$ 
is the same as in Fig. 1. }} 
\label{figtwo}
\end{figure}

\begin{figure}
\centerline{
\epsfxsize=3.5in
\epsfbox{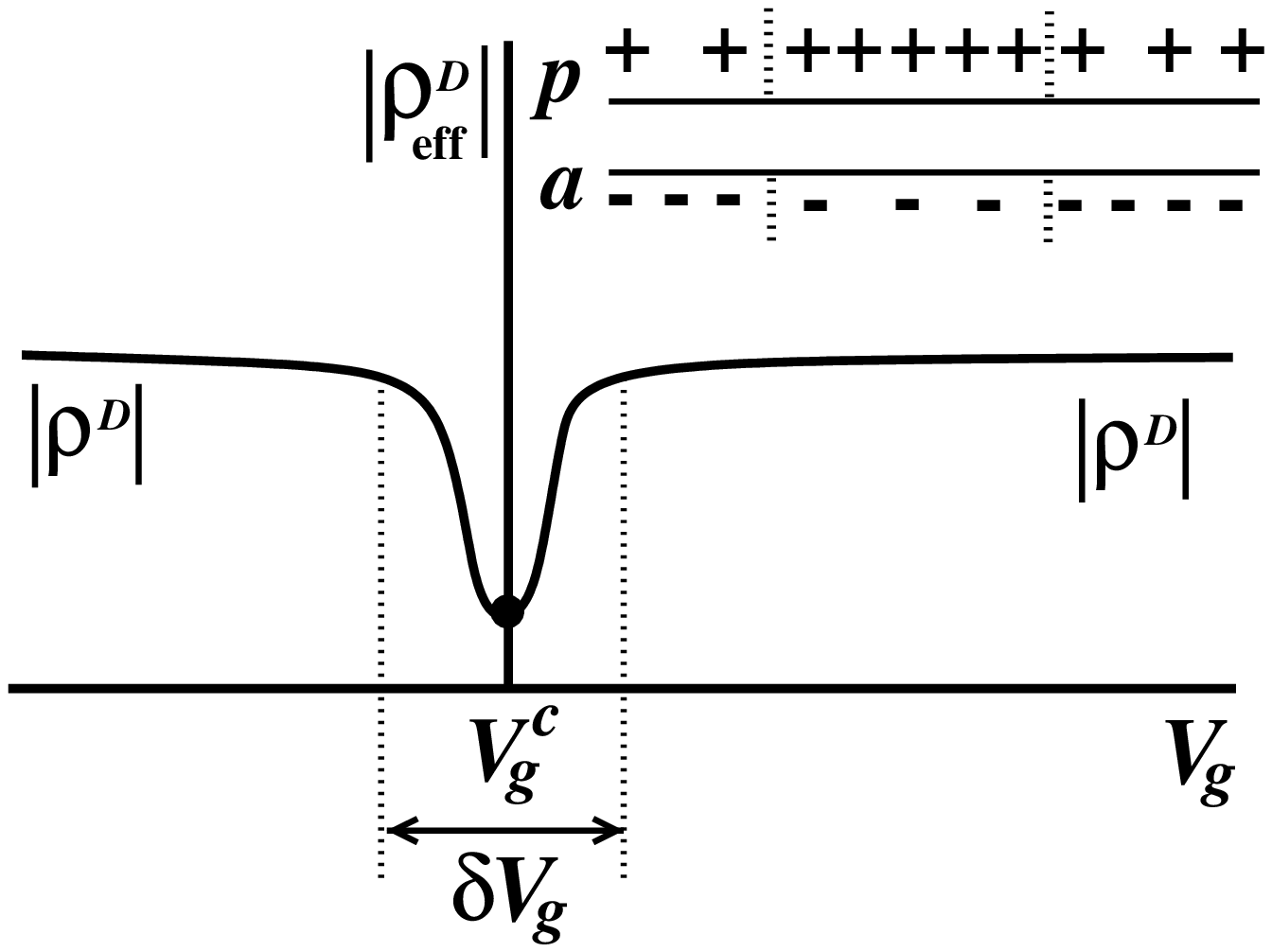}
}
\vspace*{0.1in}
\protect\caption[sample]
{\sloppy{The transresistance across the MIT is depicted schematically
for  electron-hole  system. The value
$\rho^{\mbox{\tiny $D$}}_{\mbox{\small eff}}$ at MIT is given by
Eq.  (\ref{case3_1}). The dependence $\rho^a_{\mbox{\small eff}}(V_g)$ 
is the same as in Fig. 1.}}
\label{figthree}
\end{figure}


\begin{references}

%Mutual drag of carriers in a semiconductor-insulator-semiconductor
%system
\bibitem{pogrebinskii77}M. B. Pogrebinskii, Fiz. Tekh. Poluprovodn.
{\bf 11}, 637 (1977) [Sov. Phys. Semicond. {\bf 11}, 372 (1977)].


%Hot electron effects in heterolayers
\bibitem{price83}P. M. Price, Physica  (Amsterdam) {\bf 117B},
750 (1983).

%Mutual friction between parallel two-dimensional electron systems
\bibitem{gramila91}T. G. Gramila, J. P. Eisenstein, A. H. MacDonald,
L. N. Pfeiffer, and  K. W. West, Phys. Rev. Lett. {\bf 66},
1216 (1991).
%\bibitem{gramila91}T. G. Gramila {\it et al.},
%Phys. Rev. Lett. {\bf 66}, 1216 (1991).

%Coupled electron-hole transport
\bibitem{sivan92}U. Sivan, P. M. Solomon, and H. Shtrikman,
Phys. Rev. Lett. {\bf 68}, 1196 (1992).

%Coulomb Coupling Between Spatially Separated Electron and
%Hole Layers: Generalized Random Phase Approximation
\bibitem{tso93}H. C. Tso, P. Vasilopoulos, and F. M. Peeters,
Phys. Rev. Lett. {\bf 70}, 2146 (1993).

%Quantum Hall Effect in Coulomb Drag: Interlayer Friction in Strong
%Magnetic Fields
\bibitem{shimshoni94}E. Shimshoni and S. L. Sondhi, Phys. Rev. B
{\bf 49}, 11484 (1994).

%Coupled Electron-Hole Transport: Beyond the Mean Field Approximation
\bibitem{swierkowski95}L. \'{S}wierkowski, J. Szyma\'{n}ski, and
Z. W. Gortel, Phys. Rev. Lett. {\bf 74}, 3245 (1995).

%Drag in Paired electron-Hole layers
\bibitem{vignale96}G. Vignale and A. H. MacDonald, Phys. Rev. Lett.
{\bf 76}, 2786 (1996).

%Magneto-Coulomb drag: interplay of electron-electron interactions
%and Landau quantization
\bibitem{bonsager96}M. C. B{\o}nsager, K. Flensberg, B. Y.-K. Hu,
and A.-P. Jauho, Phys. Rev. Lett. {\bf 77}, 1366 (1996).
%Phys. Rev. B {\bf 56}, 10314 (1997).

%Coulomb drag in compressible quantum Hall states
\bibitem{ussishkin97}I. Ussishkin and A. Stern, Phys. Rev. B {\bf 56},
4013 (1997).

%Coulomb drag in double-layer systems at even-denominator filling
%factors
\bibitem{sakhi97} A. Sakhi, Phys. Rev. B {\bf 56}, 4098 (1997).

%Coulomb drag at the onset of Anderson insulators
\bibitem{shimshoni97}E. Shimshoni, Phys. Rev. B {\bf 56},
13301 (1997).

%Coulomb Drag in Systems with Tunneling Bridges
\bibitem{oreg98}Y. Oreg and A. Kamenev, Phys. Rev. Lett. {\bf 80},
2421 (1998).

%Coulomb Drag in Double Layers with Correlated Disorder
\bibitem{gornyi99}I. V. Gornyi, A. G. Yashenkin, and D. V. Khveshchenko,
Phys. Rev. Lett. {\bf 83}, 152 (1999).

%Coulomb drag in intermediate magnetic fields
\bibitem{khaetskii98}A. V. Khaetskii and Yu. V. Nazarov,
Phys. Rev. B {\bf 59}, 7551 (1999).

%Gauge drag between half-filled Landau levels
\bibitem{kim99}Y.-B Kim and A. J. Millis, Physica E {\bf 4}, 171 (1999).

\bibitem{narozhny00}B. N. Narozhny, I. L. Aleiner,
Phys. Rev. Lett. {\bf 84}, 5383 (2000).

%Mesoscopic Fluctuations of the Coulomb Drag at $\nu=1/2$
\bibitem{narozhny01}B. N. Narozhny, I. L. Aleiner, and A. Stern,
Phys. Rev. Lett. {\bf 86}, 3610 (2001).

%Oscillating Sign of Drag in High Landau Levels
\bibitem{vonOppen01}F. von Oppen, S. H. Simon, and A. Stern,
Phys. Rev. Lett. {\bf 87}, 106803 (2001).

\bibitem{felix} M. E. Raikh and F. von Oppen,
Phys. Rev. Lett. {\bf 89}, 106601 (2002).

%Direct Coulomb and Phonon-Mediated Coupling Between Spatially
%Separated Electron Gases
\bibitem{tso92}H. C. Tso, P. Vasilopoulos, and F. M. Peeters,
Phys. Rev. Lett. {\bf 68}, 2516 (1992).

%Coulomb drag between parallel two-dimensional electron systems
\bibitem{jauho93}A.-P. Jauho and H. Smith, Phys. Rev. B {\bf 47},
4420 (1993).

%Coulomb drag between disordered two-dimensional electron gas layers
\bibitem{zheng93}L. Zheng and A. H. MacDonald, Phys. Rev. B
{\bf 48}, 8203 (1993).

%Coulomb drag as a probe of coupled plasmon modes in parallel
%quantum wells
\bibitem{flensberg94}K. Flensberg and B. Y.-K. Hu, Phys. Rev. Lett.
{\bf 73}, 3572 (1994).

%Coulomb drag in normal metals and superconductors: Diagrammatic
%approach
\bibitem{kamenev95}A. Kamenev and Y. Oreg, Phys. Rev. B {\bf 52},
7516 (1995).

%Linear response theory of Coulomb drag in coupled electron systems
\bibitem{flensberg95}K. Flensberg, B. Y.-K. Hu, A.-P. Jauho,
and J. Kinaret, Phys. Rev. B {\bf 52}, 14761 (1995).
%\bibitem{flensberg95}K. Flensberg {\it et al.}, Phys. Rev. B {\bf 52},
%  14761 (1995).

%Plasmon enhancement of Coulomb drag in double-quantum-well systems
\bibitem{flensberg'95}K. Flensberg, B. Y.-K. Hu,
Phys. Rev. B {\bf 52}, 14796 (1995).

\bibitem{simon94} S. H. Simon and B. I. Halperin, Phys. Rev. Lett.
{\bf 73}, 3278 (1994).

\bibitem{ruzin96} I. M. Ruzin, N. R. Cooper, and B. I. Halperin
       Phys. Rev. B {\bf 53}, 1558 (1996). 

\bibitem{stauffer92} D. Stauffer and A. Aharony, 
{\em Introduction to Percolation Theory}
 (Taylor and Francis, London, 1992).

\bibitem{dykhne1}A. M. Dykhne, Zh. Eksp. Teor. Fiz. {\bf 59}, 110 (1970)
 [Sov. Phys. JETP {\bf 32}, 3263 (1971)].

\end{references}
\end{document}